# Effect of carrier recombination on ultrafast carrier dynamics in thin films of the topological insulator Bi$_2$Se$_3$

Yuri D. Glinka,[1,2]* Sercan Babakiray,[1] Trent A. Johnson,[1] Mikel B. Holcomb,[1] and David Lederman[1]

[1]*Department of Physics and Astronomy, West Virginia University, Morgantown, WV 26506-6315, USA*
[2]*Institute of Physics, National Academy of Sciences of Ukraine, Kiev 03028, Ukraine*

Transient reflectivity (TR) from thin films (6 - 40 nm thick) of the topological insulator Bi$_2$Se$_3$ reveal ultrafast carrier dynamics, which suggest the existence of both radiative and non-radiative recombination between electrons residing in the upper cone of initially unoccupied high energy Dirac surface states (SS) and holes residing in the lower cone of occupied low energy Dirac SS. The modeling of measured TR traces allowed us to conclude that recombination is induced by the depletion of bulk electrons in films below ~20 nm thick due to the charge captured on the surface defects. We predict that such recombination processes can be observed using time-resolved photoluminescence techniques.

Thin films of topological insulators (TIs) are three-dimensional (3D) materials that are insulating in the bulk (bandgap of Bi$_2$Se$_3$, for example, $E_g$ ~ 0.3 eV), but conductive at the surfaces due to two-dimensional (2D) Dirac surface states (SS) caused by the combination of strong spin-orbit interaction and time-reversal symmetry.[1,2] Ultrafast properties of TI's remain an important issue for potential applications of these unique materials in novel electronic and optoelectronic devices. The ultrafast carrier dynamics in TIs have been studied using a variety of techniques such as time-resolved angle-resolved photoemission spectroscopy (TrARPES),[3-7] transient reflectivity (TR),[8-10] and transient second harmonic generation (TSHG).[11] The main trends elucidated include (i) the rapid electron-electron and electron-longitudinal-optical(LO)-phonon relaxation (~2-3.5 ps),[3-11] (ii) a metastable population of the 3D conduction band edge, which continuously feeds 2D Dirac SS (>10 ps),[3,10] and (iii) the quasi-equilibrium carrier population in the Dirac SS persisting within long time of about 10 ns.[10]

However, the picture emerging should be revised according to a recently discovered second set of unoccupied Dirac SS (2SS) with similar electronic structure and physical origin to the common Dirac SS (1SS) but located ~1.5 eV above the 3D conduction band edge.[7] The direct optical coupling of incident Ti:sapphire laser light, which is usually utilized to study ultrafast carrier dynamics in TIs,[8-11] to Dirac SS is expected to significantly affect the carrier relaxation dynamics since the high energy 2D Dirac 2SS overlap with the high energy 3D bands. Moreover, because direct electron-phonon relaxation through the Dirac point of Dirac 2SS seems to be suppressed due to the vanishing density of states at the Dirac point, one can expect a relaxation bottleneck, allowing for population inversion and 2D carrier recombination lasing.[12]

In this Letter we report on TR measured from thin films of the TI Bi$_2$Se$_3$, which supports the existence of both radiative and non-radiative recombination between electrons residing in the upper cone of Dirac 2SS and holes residing in the lower cone of Dirac1SS. The modeling of measured TR traces allowed us to recognize whether given 2D carrier recombination will occur.

Experiments were performed on Bi$_2$Se$_3$ thin films of various thicknesses ($d$ = 6 - 40 nm). The films were grown on 0.5 mm Al$_2$O$_3$(0001) substrates by molecular beam epitaxy, with a 10 nm thick MgF$_2$ capping layer to protect against oxidation. The growth process was similar to that described previously.[10,13,14] The films were found to be naturally $n$-doped due to remnant Se vacancies.[8-10] The free-carrier density in the films was measured using the Hall effect,[13] and is typical for as-grown Bi$_2$Se$_3$ films and single crystals. TR measurements were performed at room temperature using a Ti:Sapphire laser with a pulse duration of $\tau_L$ = 100 fs, a center photon energy of 1.51 eV and a repetition rate of 80 MHz. A wide range of laser powers was employed for pump-probe measurements, where the $s$-polarized pump was at normal incidence and the $p$-polarized probe was at an incident angle of ~ 15°, focused through the same lens to a spot diameter of ~ 100 μm. The reflection was collected in $p$-polarization by a Si photodiode and the signal was measured with a lock-in amplifier. We note that laser heating effects can be eliminated from our consideration since no film damage was observed for the laser powers used in this study.

The TR traces show a rise and multiple decay behavior, which is typical for Bi$_2$Se$_3$ thin films (Fig. 1).[10] The traces were fitted by a multi-exponential rise-decay function to obtain the rise- and decay-time constants [Fig 1(a)].[15] The rise-time constant $\tau_R$ varies in the range of 0.4 - 0.65 ps with laser power and film thickness and is associated with carrier-carrier thermalization [Fig. 2(a)].[8-11] The decay-time constants $\tau_{D1}$ ~ 1.5 - 3.1 ps, $\tau_{D2}$ ~ 5 - 220 ps, and $\tau_{D3}$ ~ 10 ns are due to intraband electron-LO-phonon relaxation, interband relaxation involving Dirac 2SS [Fig. 2(b)] and a final localization of photoexcited carriers and their recombinations, respectively [Fig. 2(c)]. Because of the natural $n$-doping of Bi$_2$Se$_3$ films (~$10^{19}$ cm$^{-3}$) and the high density of photoexcited carriers (~$10^{20}$ cm$^{-3}$),[10] the partially relaxed electrons residing above the 3D conduction band edge are characterized by a larger wavevector than 2D holes residing in the lower cone of Dirac 1SS. To conserve the wavevector during optical transitions, recombination is strictly forbidden and requires the cooperation of phonons. However, because the highest phonon energy (23 meV)[16] is much smaller than the hot carrier energy,[10,17] the phonon-assisted recombination is also less probable. Consequently, the population of photoexcited carriers in the final localization stage decays during long time ≥10 ns. This behavior agrees with the well-known suppression of recombination rate when a high density of electron-hole pairs is excited[18] and points to a specific *in-plane* wavevector conservation for carrier recombination in Dirac SS.

_________________________________________________

*Corresponding author: ydglinka@mail.wvu.edu

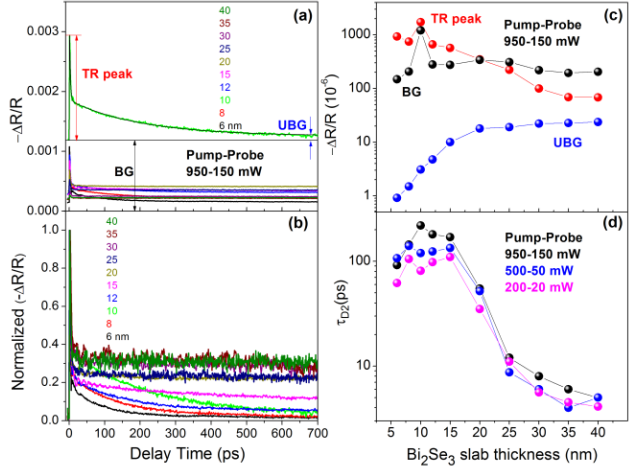

FIG. 1. (a) TR traces for $Bi_2Se_3$ films of various thicknesses indicated by the corresponding colors. An example of the fit is shown for 10 nm film together with the TR peak, UBG, and BG signal assignments. (b) Same TR traces but being normalized. (c) and (d) The thickness dependence of the amplitude of the TR peak, UBG, BG signals, and $\tau_{D2}$ constant.

Owing to the large $\tau_{D3}$ constant, there is a background (BG) signal, which is independent of the delay time between the pump and probe pulses and exists because the photoexcited carriers do not fully relax during the time between the two sequential laser pulses (12.5 ns).[10] Subsequently, the pump-induced population decays to the 3D conduction band edge and Dirac 1SS at rate of $(\tau_{D2})^{-1}$ and determines the amplitude of the unrecovered background (UBG) signal, which decays at rate of $(\tau_{D3})^{-1}$. Figure 1(c) shows that the TR peak intensity increases with decreasing $d$ according to the corresponding increase in the absorption coefficient.[10] Alternatively, the UBG signal intensity gradually decreases for films below ~20 nm thick. The BG signal intensity remains almost unchanged, excepting for a resonance-like feature for the 10 nm film, which also appears for the TR peak intensity. This behavior points to different regimes, at which the coupling between two surfaces of the film depletes 3D electrons below ~10 nm by the charge captured on the surface defects (Se vacancies), become gradually isolated with increasing film thickness and are completely decoupled beyond ~20 nm when the sum of depletion widths associated with each surface of the film is below the film thickness. The 3D-carrier-depletion-induced indirect coupling occurs because the depletion field stabilizes carriers at Dirac SS and seems to appear similarly to those observed through the compensation doping[19] and a backgate voltage.[20] We note that this effect occurs over a length scale greater than the critical thickness (~6 nm) of direct coupling.[21] It should also be stressed that the resonance-like feature at 10 nm is continuously repeatable for several films grown at the same conditions and hence indicates an additional resonant effect, the nature of which still remains unclear.

Although the $\tau_{D1}$ constant decreases with decreasing film thickness, becoming more metallic type,[10] the $\tau_{D2}$ constant increases and tends to stabilize below ~15 nm [Fig. 1(d)]. Furthermore, as it is demonstrated for the 12 nm film [Fig. 3(c)], the $\tau_{D2}$ constant remains about unchanged with increasing laser power up to the critical value, at which there is its sharp increase accompanied by a drop of the UBG signal. We initially predicted that similar behavior observed for the 10 nm thick film might be associated with the high energy saddle-point-type states formed due to direct coupling between Dirac cones at opposite surfaces of the films.[10] However, a recent ARPES-based discovery of Dirac 2SS convinced us to reassign those high energy states to Dirac 2SS.[7] Consequently, a variation of the UBG signal intensity and the $\tau_{D1}$ and $\tau_{D2}$ constants also suggests the 3D-carrier-depletion-induced suppression of the 3D LO-phonon relaxation channel, occurring for highest laser powers and film thicknesses below ~20 nm.

Because the lifetime of 3D LO phonons $\tau_{LO} = \hbar/\Delta E$,[22] where $\hbar = h/2\pi = 5.3\times10^{-12}$ cm$^{-1}$s, $h$ is Planck's constant, and $\Delta E$ is the Lorentzian linewidth of the corresponding Raman peak,[16] ($\tau_{LO}$ = 200-400 fs) is much shorter than the typical electron-LO-phonon relaxation time in $Bi_2Se_3$ (2-3.5 ps),[8-10] $\tau_{D1}$ constant is mainly governed by the sequential LO phonon Fröhlich scattering events.[10] The short LO phonon lifetime points to the highly efficient anharmonic three-phonon decay process involving acoustic phonon branches.[23,24] Consequently, the electron-LO-phonon relaxation time can be expressed as

$$\tau_{D1} = \frac{E_i \tau_{e-LO}}{h\nu_{LO}} + \tau_{LO} \approx \frac{E_i \tau_{e-LO}}{h\nu_{LO}} , \quad (1)$$

where $E_i/h\nu_{LO}$ denotes the total number of sequential electron-LO-phonon scattering events ($\tau_{e-LO}$ = 31 fs) within the LO-phonon cascade over the energy separation between electronic states $E_i$ [Fig. 3(d)] and $h\nu_{LO}$ = 8.8 meV is the LO phonon energy.[10]

One can estimate the Fermi energy[25] for free-carrier density in the 12 nm film ($n_e \sim 2.54 \times 10^{19}$ cm$^{-3}$), for example, as $E_{F1} = (\hbar^2/2m^*) \times (3\pi^2 n_e)^{2/3}$ = 242 meV, where $m^* = 0.13 m_0$ is the electron effective mass with $m_0$ being the free-electron mass.[10] Consequently, the Fermi level is energetically located above the Dirac point of Dirac 1SS, exceeding the edge of the 3D conduction band [Fig. 2(a)].[17] To figure out the spatial distribution of free carriers in the film, the depletion-layer width should be taken into account.[17,26] Correspondingly, the sum of Thomas-Fermi screening lengths associated with each surface of the film $L_{TF} = 1/q_{TF}^{top} + 1/q_{TF}^{bottom}$ can be estimated as $L_{TF}$ = 6.45+6.62=13.07 nm, where $q_{TF} \equiv \sqrt{3e^2 n_e/2\varepsilon_s \varepsilon_0 E_F}$ is the Thomas-Fermi screening wavevector,[25] $e$ is electron charge, $\varepsilon_s = \varepsilon_s^{Bi2Se3} + \varepsilon_s^{MgF2} \approx$ 113+5.3=118.3 and $\varepsilon_s = \varepsilon_s^{Bi2Se3} + \varepsilon_s^{Al2O3} \approx$ 113+11.5=124.5 are the effective low-frequency dielectric constant for the top and bottom surface of $Bi_2Se_3$ films,[17] and $\varepsilon_0$ is the permittivity of free space. The obtained value exceeds the film thickness and hence allows for intersurface coupling, efficiently forming the depletion region over the film thickness.[20] Correspondingly, the films below 15 nm thick behave like a poor metal due to the depletion of 3D electrons, giving rise to a decrease of 3D conductivity that appears in transport measurements.[17,26,27] Furthermore, such unscreened surface potentials mask 3D electron-LO-phonon interaction in the films, suppressing the LO-phonon oscillatory part of the TR signals observed with decreasing $d$.[10,28] For thicker films ($\geq$15 nm) $L_{TF}$ is always shorter than $d$, indicating that the coupling between the film surfaces is gradually decreased.

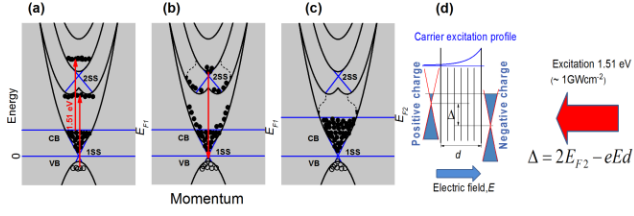

FIG. 2. Schematic presentation of the excitation with 1.51 eV photons and carrier-carrier thermalization (a), the intraband and interband electron-LO-phonon relaxation involving Dirac 2SS-1SS carrier recombination (b), and the final carrier localization at the 3D conduction band edge and 2D Dirac 1SS (c). The initial Fermi energy (free carriers) $E_{F1}$ and the final Fermi energy (photoexcited carriers) $E_{F2}$ are shown. (d) Schematic representation of the carrier excitation profile and the final stage of carrier localization at Dirac 1SS. The shift of Dirac points ($\Delta$) at the opposite surfaces is shown.

Once the optical excitation with photon energy of 1.51 eV is applied, the carriers can be excited in two ways. The free electrons below the Fermi level are excited into the high energy 3D band energetically located above the Dirac point of Dirac 2SS. The common valence-to-conduction band transitions concurrently excite also 3D electrons and holes [Fig. 2(a)]. For the 12 nm film, for example, and for the highest average laser power applied 1100 mW (corresponds to 1.76 GW/cm$^2$ power density) one can obtain the photoexcited carrier density $n \sim 1.47 \times 10^{20}$ cm$^{-3}$ and $n \sim 5.4 \times 10^{19}$ cm$^{-3}$ for the top and bottom surface of the film.[10] Here we assumed that the carrier excitation profile in the film follows the Beer's law and hence the density of photoexcited carrier on the top surface of $Bi_2Se_3$ films is about number-$e$ times larger than that on the bottom surface [Fig. 2(d)]. Consequently, the sum of Thomas-Fermi screening lengths for photoexcited carrier densities can be estimated as $L_{TF}$ = 4.81+5.83=10.64 nm, which is shorter than $d$ and therefore the opposite surfaces of the film dynamically become loosely coupled. The corresponding Fermi energies can be estimated as $E_{F2}$ = 780 meV and $E_{F2}$ = 400 meV for the top and bottom surfaces of the film, respectively. As a result of Fermi energy difference ($\Delta$) at the opposite surfaces, the 3D-carrier-depletion-mediated capacitor-like electric field directed along the film normal is expected to form. This charged quasi-stationary condition can exist during long time, corresponding to the UBG signal. Once the system reaches quasi-equilibrium, the capacitor-like electric field $E = (2E_{F2} - \Delta)/ed$ gradually balances the top and bottom surface Fermi energies by shifting the opposite surface Dirac points of Dirac 1SS [Fig. 2(d)].[29]

However, this scenario can be significantly modified by recombination between electrons and holes, efficiently decreasing the population of photoexcited carriers and hence the amplitude of UBG signal observed with decreasing $d$ [Fig. 1(c)] and with increasing laser power [Fig. 3(c)]. To explain the ultrafast carrier dynamics we consider a model, which uses the following assumptions [Fig. 3(d)]: (i) the effective time of LO-phonon relaxation is governed by the total number of sequential electron-LO-phonon scattering events [Eq. (1)]; (ii) LO-phonon relaxation occurs through all the 3D electronic states; (iii) the radiative and non-radiative recombination occurs between electrons residing in the upper cone of Dirac 2SS and holes residing in the lower cone of Dirac 1SS; (iv) the carrier recombination takes much longer time than LO-phonon relaxation because it requires a spin flip;[30] (v) the electron-phonon relaxation through the Dirac point of Dirac 2SS is suppressed due to the vanishing density of states at the Dirac point; and (vi) the partial accumulation of electrons in the upper cone of Dirac 2SS with close-to-zero wavevectors is mediated by the depletion of 3D electrons in the film, allowing the *in-plane* wavevector to be conserved in recombination with holes in the lower cone of Dirac 1SS.

Using this model, the population of photoexcited electrons, which thermalizes due to the electron-electron scattering, obeys the following rate equation:

$$\partial n/\partial t = G - (\tau_{e-e})^{-1} n , \quad (2)$$

where $G$ is the photoexcited electrons rate and $(\tau_{e-e})^{-1}$ is the electron-electron scattering rate [~ $(\tau_R)^{-1}$]. Subsequently, the density of thermalized electrons increases with time as

$$n_0 = [G/(\tau_{e-e})^{-1}] \times [1 - e^{-(\tau_{e-e})^{-1} t}] . \quad (3)$$

The density of thermalized holes ($p_0$) due to hole-hole scattering can be expressed analogously. Both densities define the initial density of photoexcited carriers in the system since the laser-pulse-excited electron-hole plasma is neutral and hence it does not contribute to the TR dynamics.

As a consequence of electron-(hole)-LO-phonon relaxation, the thermalized electron (hole) population decays to the edge of the high energy 3D band, defining $n_1$ (to the edge of the 3D valence band and subsequently to the lower cone of 2D Dirac 1SS, defining $p_{1SS}$). Assuming that ultrafast carrier dynamics are mainly governed by electrons while photoexcited holes only contribute into recombination, the subsequent decay of photoexcited electrons occurs via two channels. The first channel includes the 3D LO-phonon interband relaxation to the lower energy bands [Fig. 3(d)]:

$$\partial n_1/\partial t = n_0 \frac{h\nu_{LO}}{E_1}(\tau_{e-LO})^{-1} - n_1 \frac{h\nu_{LO}}{E_2}(\tau_{e-LO})^{-1} , \quad (4)$$

where $h\nu_{LO}(\tau_{e-LO})^{-1}/E_i$ is the rate of electron-LO-phonon scattering [Fig. 3(d)]. The solution of Eq. (4) is

$$n_1 = n_0(E_2/E_1) \times \left[1 - e^{-\frac{h\nu_{LO}}{E_2}(\tau_{e-LO})^{-1} t}\right]. \quad (5)$$

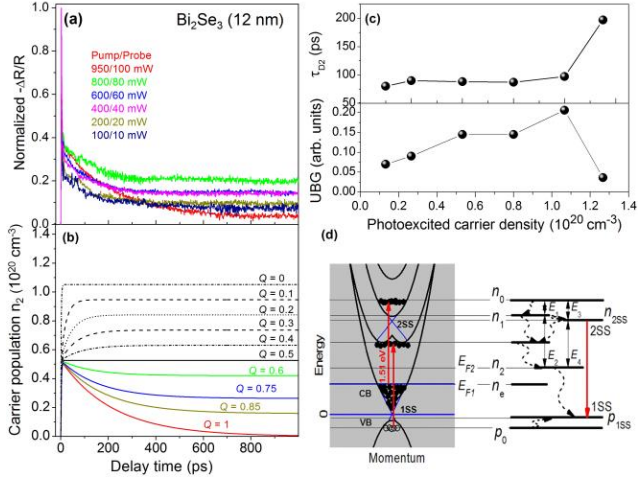

FIG. 3. (a) Normalized TR traces for the 12 nm $Bi_2Se_3$ film measured with different pump/probe power indicated by the corresponding colors. (b) Numerical calculations of photoexcited carrier density ($n_2$) using Eq. (11) and $n_0 \sim 1.05 \times 10^{20}$ cm$^{-3}$ (800 mW), $\tau_{e-e} = 0.5$ ps, $\tau_{D1} = 2$ ps, $\tau = 220$ ps, $\tau_{1SS} = 10$ ns. (c) The power dependence of the $\tau_{D2}$ constant and the UBG signal. (d) The schematic band diagram of the TI $Bi_2Se_3$ showing the initial stage of carrier excitation (left) and the corresponding energy level diagram showing the carrier relaxation channels and recombination transitions (right).

The second channel involves the population and depopulation of the upper cone of Dirac 2SS,

$$\frac{\partial n_{2SS}}{\partial t} = n_0 \frac{h\nu_{LO}}{E_3}(\tau_{e-LO})^{-1} - n_{2SS}\left[\frac{h\nu_{LO}}{E_4}(\tau_{e-LO})^{-1} + \tau^{-1}\right], \quad (6)$$

where $\tau^{-1} = (\tau_R)^{-1} + (\tau_{NR})^{-1}$ is the total rate of radiative and non-radiative recombination [$\sim (\tau_{D2})^{-1}$].

The solution of Eq. (6) is

$$n_{2SS} = n_0 Q \frac{h\nu_{LO}(\tau_{e-LO})^{-1}}{E_3 \tau^{-1}}\left[1 - e^{-\frac{\tau^{-1}}{Q}t}\right], \quad (7)$$

where

$$Q = \left[1 + \frac{h\nu_{LO}(\tau_{e-LO})^{-1}}{E_4 \tau^{-1}}\right]^{-1} \quad (8)$$

is a parameter that controls the second channel efficiency and is governed by the ratio between the rates of 3D LO-phonon carrier relaxation and carrier recombination through 2D Dirac SS. Finally, the population of the 3D conduction band edge, which governs the amplitude of the UBG signal, varies with time as

$$\frac{\partial n_2}{\partial t} = n_1 \frac{h\nu_{LO}}{E_2}(\tau_{e-LO})^{-1} + n_{2SS}\left[\frac{h\nu_{LO}}{E_4}(\tau_{e-LO})^{-1} - \tau^{-1}\right] - n_2(\tau_{1SS})^{-1}, \quad (9)$$

where $(\tau_{1SS})^{-1}$ is the rate of both radiative and non-radiative relaxation in Dirac 1SS [$\sim (\tau_{D3})^{-1}$].

Because $\tau_{D3}$ is $\geq 10$ ns, Eq. (9) can be solved using a quasi-stationary condition ($\partial n_2/\partial t = 0$):

$$n_2 = n_1\left[\frac{h\nu_{LO}}{E_2}\frac{(\tau_{e-LO})^{-1}}{(\tau_{1SS})^{-1}}\right] + n_{2SS}\left[\frac{h\nu_{LO}}{E_4}\frac{(\tau_{e-LO})^{-1}}{(\tau_{1SS})^{-1}} - \frac{\tau^{-1}}{(\tau_{1SS})^{-1}}\right]. \quad (10)$$

Finally, the overall dynamics yields the solution

$$n_2 = An_0 \times \left\{\frac{E_2}{E_1}\left[1 - e^{-A(\tau_{1SS})^{-1} t}\right] + (1-2Q)\frac{E_2}{E_3}\left[1 - e^{-\frac{\tau^{-1}}{Q}t}\right]\right\}, \quad (11)$$

with $A = h\nu_{LO}(\tau_{e-LO})^{-1}/E_2(\tau_{1SS})^{-1}$. Equation (11) has a simple physical meaning that the photoexcited electrons tend to reside above the edge of the 3D conduction band at rate corresponding to 3D LO-phonon relaxation and the final electron density (and $E_{F2}$) is completely determined by $Q$.

Figure 3(b) shows that by varying parameter $Q$ the long delay-time ($> 3$ ps) dynamics of measured TR traces can be well reproduced. The traces shown in Figure 1 clearly demonstrate that the 2D carrier recombination rate can be significantly enhanced in films below $\sim$20 nm thick. We associate this behavior with the depletion of 3D electrons and the resulting stabilization of 2D electrons at the Dirac SS. Because the rate of carrier relaxation in Dirac 1SS ($\tau_{1SS}$)$^{-1}$ is extremely small, one can observe an increase of the UBG signal intensity with increasing laser power followed by a drop when the 2D carrier recombination rate becomes dominant [Fig. 3(c)]. Furthermore, the sharp increase of $\tau_{D2}$ constant observed for highest laser powers [Fig. 3(c)] suggest the comparable rates ($Q \sim 0.7$) of the 3D LO-phonon carrier relaxation and 2D carrier recombination at the moderate laser powers and their suppression and enhancement, respectively, above the critical density of photoexcited carrier ($Q \sim 1$).

In summary, our experimental results on TR from thin films of the TI $Bi_2Se_3$ suggest the existence of both radiative and non-radiative recombination between electrons residing in the upper cone of Dirac 2SS and holes residing in the lower cone of Dirac1SS. The 2D carrier recombination rate is expected to be significantly enhanced by the depletion of 3D electrons in films below $\sim$20 nm thick. We predict that such recombination transitions can be observed using time-resolved photoluminescence techniques.

This work was supported by a Research Challenge Grant from the West Virginia Higher Education Policy Commission (HEPC.dsr.12.29). Some of the work was performed using the West Virginia University Shared Research Facilities.